\documentclass[aps,prl,showpacs,twocolumn,floats]{revtex4}
\usepackage{graphicx}
\usepackage{subfigure}
\usepackage{epsfig}
\usepackage{dcolumn}
\usepackage{bm}

\begin{document}

\title{Quantum Tunneling of Normal-Superconducting Interfaces in a Type-I Superconductor}
\date{\today}
\author{E. M. Chudnovsky$^{1,2}$, S. V\'{e}lez$^{2,3}$, A. Garc\'{i}a-Santiago$^{2,3}$, J. M. Hernandez$^{2,3}$, and J. Tejada$^{2,3}$}
\affiliation{$^{1}$Physics Department, Lehman College, The City
University of New York, 250 Bedford Park Boulevard West, Bronx, NY
10468-1589, U.S.A.\\ $^{2}$Departament de F\'{i}sica Fonamental,
Facultat de F\'{i}sica, Universitat de Barcelona, Avinguda
Diagonal 645, 08028 Barcelona, Spain\\ $^3$Institut de
Nanoci\`{e}ncia i Nanotecnologia IN2UB, Universitat de Barcelona,
c. Mart\'{i} i Franqu\`{e}s 1, 08028 Barcelona, Spain}
\date{\today}

\begin{abstract}
Evidence of a non-thermal magnetic relaxation in the intermediate
state of a type-I superconducor is presented. It is attributed to
quantum tunneling of interfaces separating normal and
superconducting regions. Tunneling barriers are estimated and
temperature of the crossover from thermal to quantum regime is
obtained from Caldeira-Leggett theory. Comparison between theory
and experiment points to tunneling of interface segments of size
comparable to the coherence length, by steps of order one
nanometer.
\end{abstract}

 \maketitle

Quantum tunneling of relatively macroscopic solid-state objects
like flux lines in type-II superconductors
\cite{Larkin-review1994,Nattermann-review2000} and domain walls in
magnets \cite{CT-book} have been subject of intensive research in
the past. Pinning by defects and impurities creates complex
potential landscape that traps flux lines and domain walls inside
metastable energy minima. Their interaction with environment makes
this problem the one of macroscopic quantum tunneling with
dissipation \cite{Caldeira-Leggett}. The latter is especially
important for the tunneling of flux lines because of their
predominantly dissipative dynamics
\cite{Blatter-1991,Ivlev-1991,Tejada-1993,Ao-1994,Stephen-1994}.
If individual pinning centers are small compared to the dimensions
of the vortex or the width of the domain wall, the pinning is
collective. In this case the energy barriers and spatial scales of
thermal and quantum diffusion of flux lines and domain walls are
non-trivially determined by statistical mechanics of the pinning
potential
\cite{Larkin-review1994,Nattermann-review2000,Nattermann-1990,Giamarchi-PRB2000}.

\begin{figure}
\vspace{-3mm}
\includegraphics[width=79mm]{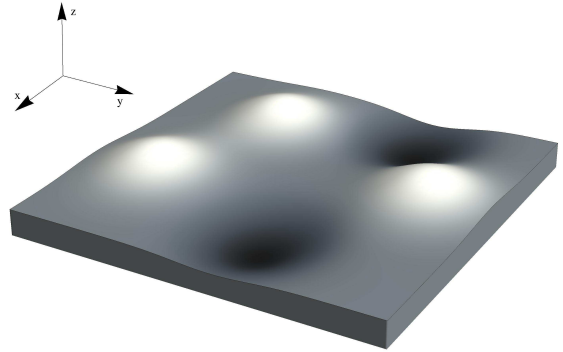}
\caption{Color online: Schematic view of the interface between
normal and superconducting regions of type-I superconductor,
slightly perturbed by randomly scattered pinning centers.}
\label{interface}
\end{figure}

When placed in the magnetic field, type-I superconductors do not
develop flux lines. Instead, they exhibit intermediate state in
which the sample splits into normal and superconducting regions
separated by planar interfaces of positive energy
\cite{Landau,Sharvin,Huebener}. Recently, there has been a renewed
interest to the equilibrium structure, pinning, and dynamics of
interfaces in type-I superconductors
\cite{Kuznetsov-1998,Cebers-2005,Menghini-2005,Prozorov-PRL2007,Prozorov-Nature2008,Velez-PRB2009}.
Pure lead has been mostly used as a prototypical experimental
system. In the presence of pinning centers the interfaces adjust
to the pinning potential by developing curvature as is
schematically shown in Fig.\ \ref{interface}. Pinning by point or
small-volume defects should result in a broad distribution of
energy barriers. It is, therefore, plausible that at low
temperature type-I superconductors continue to relax towards
equilibrium via quantum diffusion of interfaces. This situation is
similar to the diffusion of domain walls in disordered
ferromagnets with one essential difference. Contrary to a
ferromagnetic domain wall, the dynamics of the planar interface in
a superconductor should be dominated by dissipation.

At low temperature the decay of metastable states created by
pinning provides slow relaxation of magnets and superconductors
towards thermal equilibrium. This relaxation is known as magnetic
after-effect. At finite temperature it may occur via thermal
activation with a probability proportional to $\exp(-U_B/T)$ where
$U_B$ is the energy barrier. As $T \rightarrow 0$ thermal
processes die out and the only channel of escape from the
metastable state becomes underbarrier quantum tunneling. Its
probability is proportional to $\exp(-I_{eff}/\hbar)$ where
$I_{eff}$ is the effective action associated with tunneling. The
pre-exponential factors in the two expressions are of lesser
importance because the dependence of the probability on the
parameters is dominated by the exponents. Equating the two
exponents, one finds that the crossover from thermal activation to
quantum tunneling occurs at $T_Q \approx {\hbar U_B}/{I_{eff}}$.
Experimental evidence of such a crossover in type-II
superconductors has been overwhelming \cite{Yeshurun-review1996}.
There has also been some experimental evidence of quantum
diffusion of domain walls in disordered ferromagnets
\cite{Rosenbaum}. However, to our knowledge, no literature exists
on non-thermal magnetic relaxation in type-I superconducors.
Experimental evidence of such a relaxation and its theoretical
treatment are subjects of this Letter.

The lead sample used in our experiments was an octagonal disk of
thickness $0.2\,$mm and surface area $40\,$mm$^2$, cut from a
commercial Pb rod of purity $99.999\%$. It was annealed during one
hour at $290^{\rm{o}}$C in glycerol and nitrogen atmosphere to
reduce the mechanical stress from defects that might have been
introduced during preparation of the sample. Magnetic measurements
were performed with the use of a commercial superconducting
quantum interference device (SQUID) magnetometer in the field up
to $1\,$kOe in the temperature range $2$K - $8$K with thermal
stability better than 0.01 K. Isothermal magnetization curves were
measured to obtain the temperature dependence of the thermodynamic
critical field. The fit of the data by $B_c(T) = B_{c}(0)[1
-(T/T_c)^2]$ produced $B_c(0) = 802\pm 2\,$Oe and $T_c = 7.23 \pm
0.02\,$K, in accordance with the values of the critical field and
transition temperature reported for lead.
\begin{figure}
\includegraphics[width=85mm]{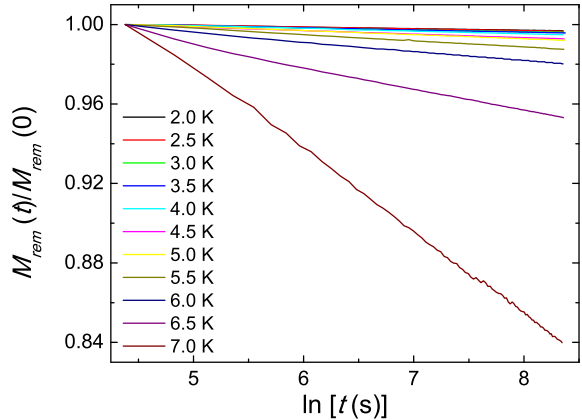}
\caption{Color online: Magnetic relaxation at various
temperatures. Logarithmic time dependence provides an accurate fit
to the data.} \label{M(t)}
\end{figure}
\begin{figure}
\includegraphics[width=85mm]{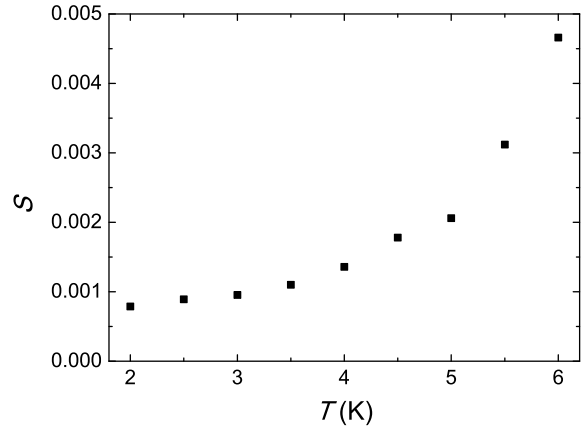}
\caption{Temperature dependence of the magnetic viscosity. $S(T)$
tends to a non-zero temperature-independent value in the limit of
$T \rightarrow 0$.} \label{viscosity}
\end{figure}

Magnetic relaxation was measured by first applying the field $B
> B_c(T)$, then subsequently switching the field off and
recording isothermal temporal evolution of the remnant
magnetization $M_{\rm{rem}}(T)$ in a zero field. Fig.\ \ref{M(t)}
shows the time evolution of $M_{\rm{rem}}(t)/M_{\rm{rem}}(0)$
between $2.00\,$K and $7.00\,$K in steps of $0.50 \,$K. At all
temperatures the observed slow relaxation followed very well the
logarithmic time dependence,
$M_{\rm{rem}}(t)=M_{\rm{rem}}(0)[1-S(T)\ln t]$, where $S(T)$ is
the so-called magnetic viscosity. Temperature dependence of $S(T)$
is shown in Fig.\ \ref{viscosity}. Remarkably it does not
extrapolate to zero in the limit of $T \rightarrow 0$ but,
instead, tends to a finite temperature-independent limit as the
sample is cooled down.

\begin{figure}
\includegraphics[width=80mm]{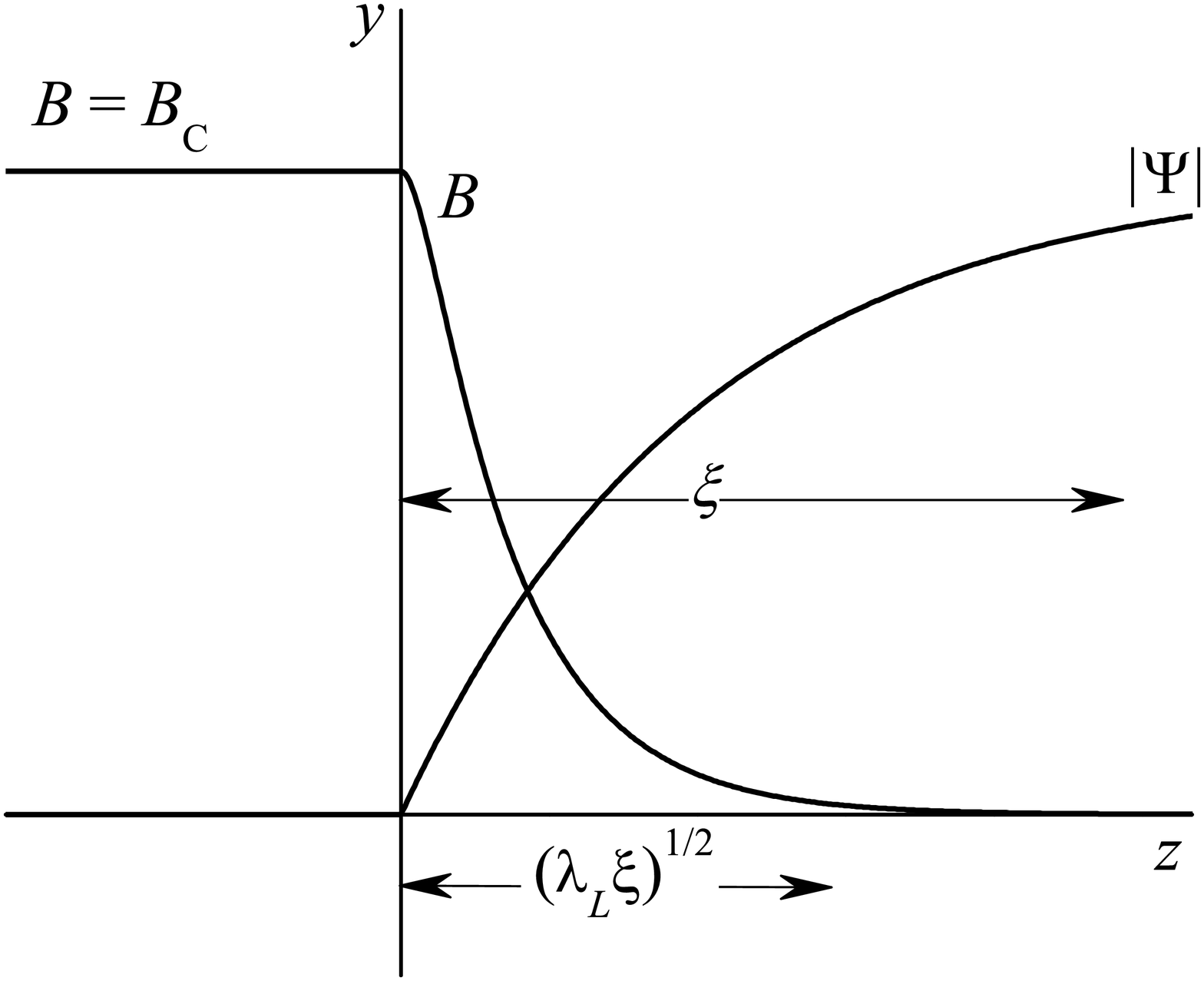}
\caption{Structure of the interface between normal and
superconducting regions of type-I superconductor. The magnetic
field decays on a scale $\delta = \sqrt{\lambda_L\xi}$, while the
modulus of the Cooper-pair condensate wave function changes on a
scale $\xi$.} \label{structure}
\end{figure}

As is well known, logarithmic time-dependence of the relaxation is
an indication of the broad distribution of the energy barriers for
the escape from metastable states \cite{CT-book}. The finite value
of $S(0)$ points towards quantum mechanism of the escape. By
analogy with type-II superconductors, where non-thermal magnetic
relaxation is due to quantum tunneling of flux lines, it is
reasonable to assume that in type-I superconductors the effect is
due to quantum tunneling of interfaces separating normal and
superconducting regions.  The structure of the interface (see
Fig.\ \ref{structure}) is determined by two parameters: the
coherence length $\xi$ and the London length $\lambda_L$. Type-I
superconductivity corresponds to $\kappa = \lambda_L/\xi <
1/\sqrt{2}$. Concentration of Cooper pairs $|\Psi|^2$ gradually
goes to zero on a distance $\xi$ as one moves through the
interface from the superconducting to the normal region. When
crossing the interface in the opposite direction one would see the
magnetic field going down from its thermodynamic critical value
$B_c$ to zero on a distance $\delta = \sqrt{\lambda_L\xi} < \xi$.

The energy of the unit area of the interface is \cite{LP}
\begin{equation}\label{sigma}
\sigma = \frac{\xi B_c^2}{3\sqrt{2}\pi}\,.
\end{equation}
Pinning provides curvature of the interface, see Fig.\
\ref{interface}. We shall describe such an interface by a
singled-valued function $Z(x,y,\tau)$. The potential energy of the
interface consists of two parts:
\begin{eqnarray}\label{energy}
E & = & \sigma \int dx dy \left[1 + \left(\frac{dZ}{dx}\right)^2 +
\left(\frac{dZ}{dy}\right)^2\right]^{1/2} \nonumber \\
& + & \int dx dy \, U\left[x,y,Z(x,y,\tau)\right]\,.
\end{eqnarray}
The first integral is the energy of a two-dimensional elastic
manifold. The second integral is the energy due to the pinning
potential $U(x,y,z)$. Metastable equilibrium is achieved through
balance of the elastic energy and pinning energy that corresponds
to the minimum of Eq.\ (\ref{energy}). We shall assume that
magnetic relaxation occurs due to the decay (or formation) of the
bumps in the interface shown in Fig.\ \ref{interface}. We shall
describe such a bump by the lateral size $L$ and height $a$. For a
particular bump these parameters are determined by the local
pinning potential. Since the latter is unknown we shall test
self-consistency of the approach based upon theory of tunneling
with dissipation \cite{Caldeira-Leggett} by extracting the average
values of $L$ and $a$ from experiment.

\begin{figure}
\includegraphics[width=78mm]{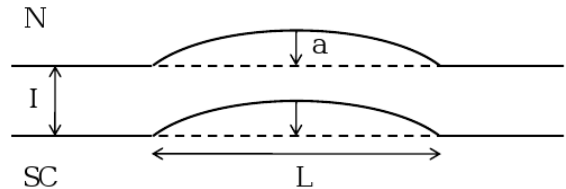}
\caption{Flattening (or formation) of a bump via quantum tunneling
of a pinned interface (I) separating normal (N) and
superconducting (SC) regions.} \label{segment}
\end{figure}

Let us first estimate the energy barrier associated with the bump.
It is easy to see that the change in the elastic energy of the
interface due to formation of the bump (see Fig.\ \ref{segment})
is independent of $L$ and is generally of order $\sigma \pi a^2$.
(This follows from the fact that the area of a spherical segment
above any cross-section of a sphere differs from the area of that
cross-section by $\pi a^2$). This energy must be balanced by the
negative energy of the pinning to make the bump an equilibrium
state of the interface. Consequently,
\begin{equation}\label{barrier}
U_B \approx \pi \sigma a^2
\end{equation}
with the average value of $a$ should represent the typical
amplitude of the random pinning potential.

We want to find the WKB exponent, $I_{eff}/\hbar$, for the
tunneling of $Z(x,y)$ between two configurations of the interface
corresponding to the local energy minima (see Fig.\
\ref{segment}).
Same as for the flux lines
\cite{Blatter-1991,Ivlev-1991,Tejada-1993,Ao-1994,Stephen-1994} we
shall assume that the tunneling probability is dominated by the
dissipation part of the Caldeira-Leggett effective action
\cite{Caldeira-Leggett}:
\begin{eqnarray}\label{I-dis}
I_{eff}& = & \frac{\eta}{4\pi}\int^{\hbar/T}_0  d\tau
\int_{-\infty}^{\infty} d\tau'
\nonumber \\
& \times & \int dx dy\, \frac{\left[Z(x,y,\tau) -
Z(x,y,\tau')\right]^2}{(\tau - \tau')^2}\,,
\end{eqnarray}
where $\eta$ is a viscous drag coefficient describing dissipative
motion of the interface and $\tau=it$ is imaginary time. For a
segment of the interface of size $L$, that tunnels by a distance
$a$, the $T = 0$ value of the effective action in Eq.\
(\ref{I-dis}) can be estimated as
\begin{equation}\label{action}
I_{eff} \approx \frac{\eta L^2 a^2}{4\pi}\,.
\end{equation}

The drag coefficient $\eta$ can be obtained from the argument
similar to that of Bardeen and Stephen for the flux lines
\cite{Bardeen}. Let magnetic field be in the $y$-direction. In the
presence of the current of density $j$ in the $x$-direction, the
magnetic force experienced by the $dxdy$ element of the interface
in the $z$-direction is
\begin{equation}
dF = \frac{1}{c}\int dx dy dz\,j B\,.
\end{equation}
Writing $j$ via the electric field and normal-state resistivity
$\rho_n$ as $j = E/\rho_n$, and substituting here $E = (V/c)B$ for
the electric field produced inside the interface moving at a speed
$V$ in the $z$-direction, one has $j = (V/c)(B/\rho_n)$. This
gives
\begin{equation}
\frac{dF}{dxdy}= \frac{V}{\rho_n c^2}\int dz B^2(z)
\end{equation}
for the force per unit area of the interface. Substitution into
this formula of $B \approx B_c\exp(-z/\delta)$ finally yields
\begin{equation}\label{eta}
\frac{dF}{dxdy}= \eta V\,, \qquad \eta = \frac{\sqrt{\lambda_L
\xi}B_c^2}{2\rho_n c^2}\,.
\end{equation}

As has been explained in the introduction, the crossover from
thermal to quantum diffusion of the interface should occur around
$T_Q = \hbar U_B/I_{eff}$. With the help of Eqs.\ (\ref{barrier}),
(\ref{action}), and (\ref{eta}) one obtains
\begin{equation}\label{Tc}
T_Q \approx \frac{4\pi^2 \hbar \sigma}{\eta L^2} =
\frac{4\pi\sqrt{2}\hbar \rho_n c^2}{3\sqrt{\kappa} L^2}\,.
\end{equation}
Notice that due to the dimensionality of the problem $T_Q$ does
not depend on the size of the tunneling step $a$. Recalling the
expression for $\lambda_L$ in terms of the effective mass $m$ and
concentration $n$ of the electrons,  $\lambda_L = [mc^2/(4\pi
e^2n)]^{1/2}$, and writing $\rho_n = (m\nu/e^2n) = 4\pi\nu
\lambda_L^2/c^2$ in terms of the normal electron collision
frequency $\nu$, the crossover temperature can be presented in the
form
\begin{equation}\label{Tc-nu}
T_Q \approx \frac{16\pi^2 \sqrt{2}}{3}
\kappa^{3/2}\left(\frac{\xi}{L}\right)^2\hbar\nu
\end{equation}
that shows its explicit dependence on the microscopic parameters
of the material.

Let us now compare our experimental findings with theoretical
results. The temperature of the crossover from thermal activation
to quantum tunneling can be estimated from the following argument.
At non-zero temperature the magnetic viscosity $S$ shown in Fig. 3
has contributions from both, thermal activation and quantum
tunneling, $S = S_T + S_Q$, where $S_Q = S(0)$. The parameter
$T_Q$ is defined as temperature at which the two contributions are
equal, that is, $S_T = S_Q$ and $S(T_Q) = 2S_Q$. This gives $T_Q$
in the ballpark of $4$K.

The values of $\lambda_L$ and $\xi$ in lead are $37$nm and $83$nm,
respectively, giving $\kappa = \lambda_L/\xi = 0.45$.
Thermodynamic critical field, $B_c$, is close to $800$G. For the
energy of the unit area of the interface Eq.\ (\ref{sigma}) gives
$\sigma \sim 0.4\,$erg/cm$^2$. Normal resistivity of lead at
$4\,$K is of order \cite{resistivity} $5 \times 10^{-11}$ $\Omega
\cdot$ m $\approx 5.6\times10^{-21}$s. Eq.\ (\ref{eta}) then gives
for the drag coefficient $\eta \approx 0.35\,$erg$\cdot$ s/cm$^4$.
We shall now check self-consistency of our model by computing the
average size of the tunneling segment $L$ and the tunneling step
$a$. From Eq.\ (\ref{Tc}) one obtains $L \approx 90\,$nm $\sim
\xi$, which is rather plausible. Indeed, $L \sim \xi$ describes
the segment of the interface inside which Cooper pairs are
strongly correlated and, therefore, they can collectively
participate in a coherent tunneling event. For the tunneling
transition to occur in our experimental time window of one hour,
$I_{eff}$ cannot significantly exceed $25 \hbar$. According to
Eq.\ (\ref{action}) this condition is satisfied by tunneling steps
$a$ below $1\,$nm, which is also quite plausible. According to
Eq.\ (\ref{barrier}), the typical energy barrier must be then of
order $100$K in accordance with the fact that thermal activation
dies out below $4\,$K.

In Conclusion, we have observed non-thermal magnetic relaxation in
lead that we attribute to quantum tunneling of small segments of
interfaces separating normal and superconducting regions. Theory
of such a tunneling has been developed. Comparison between theory
and experiment suggests macroscopic quantum tunneling of interface
segments comparable in size to the coherence length, by steps of
order one nanometer.

The work of E.M.C. has been supported by the grant No.
DE-FG02-93ER45487 from the U.S. Department of Energy and by
Catalan ICREA Academia. S.V. acknowledges financial support from
Ministerio de Ciencia e Innovaci\'{o}n de Espa\~{n}a. J.M.H. and
A.G.-S. thank Universitat de Barcelona for supporting their
research. J.T. acknowledges financial support from ICREA Academia.

\end{document}